\documentclass[%
 reprint,
nofootinbib,
&nobibnotes,
 amsmath,amssymb,
 aps,
]{revtex4-2}

\usepackage{graphicx}
\usepackage{xcolor}
\usepackage{dcolumn}
\usepackage{bm}
\usepackage{hyperref}
\hypersetup{
    colorlinks = true,
    allcolors = {teal}
}

\def \bal#1\eal  {\begin{align} #1 \end{align}}
\def \bga#1\ega  {\begin{gather} #1 \end{gather}}
\def\({\left(}
\def\){\right)}
\def\[{\left[}
\def\]{\right]}
\def\<{\left\langle}
\def\>{\right\rangle}
\def\d{\mathrm{d}}

\newcommand{\eim}{\end{itemize}}
\newcommand{\beq} {\begin{equation}}
\newcommand{\eeq} {\end{equation}}
\newcommand{\bc}{\begin{center}}
\newcommand{\ec}{\end{center}}

\newcommand{\nn} {\nonumber\\}

\newcommand{\pd} {\partial}

\newcommand{\di}{{\delta}}

\newcommand{\li}{{\lambda}}

\newcommand{\oi}{\omega}

\newcommand{\Gi}{\Gamma}

\newcommand{\mE}{\mathcal{E}}
\newcommand{\mM}{\mathcal{M}}
\newcommand{\mX}{\mathcal{X}}
\newcommand{\pr}{{\prime}}

\begin{document}

\title{Pole skipping in holographic theories with bosonic fields}

\author{Diandian Wang}
\author{Zi-Yue Wang}%
\affiliation{%
Department of Physics, University of California, Santa Barbara, CA 93106}%

\begin{abstract}
We study pole skipping in holographic CFTs dual to diffeomorphism invariant theories containing an arbitrary number of bosonic fields in the large $N$ limit. Defining a weight to organize the bulk equations of motion, a set of general pole-skipping conditions are derived. In particular, the frequencies simply follow from general covariance and weight matching. In the presence of higher spin fields, we find that the imaginary frequency for the highest-weight pole-skipping point equals the higher-spin Lyapunov exponent which lies outside of the chaos bound.
Without higher spin fields, we show that the energy density Green's function has its highest-weight pole skipping happening at a location related to the OTOC for arbitrary higher-derivative gravity, with a Lyapunov exponent saturating the chaos bound and a butterfly velocity matching that extracted from a shockwave calculation. We also suggest an explanation for this matching at the metric level by obtaining the on-shell shockwave solution from a regularized limit of the metric perturbation at the skipped pole. 

\end{abstract}

\maketitle


\section{\label{sec:intro}Introduction}

The out-of-time-order correlator (OTOC), an important quantity containing characteristics of chaos, can be calculated holographically in a shockwave spacetime \cite{Shenker:2013pqa,Roberts:2014isa,Roberts:2016wdl,Shenker:2013yza,Shenker:2014cwa}. For a localized perturbation to a chaotic system at temperature $T$, the OTOC between a perturbation $W$ at $x=t=0$ and a probe operator $V$ at a later time $t$ behaves as
\begin{equation}\label{eq:otoc}
\langle V(x,t) W V(x,t) W\rangle\sim1- e^{\lambda_L\left(t-t_*-|x| / v_{B}\right)},
\end{equation}
where $t_*$ is called the scrambling time. This defines the Lyapunov exponent, $\lambda_L$, and the butterfly velocity, $v_B$. For classical bulk gravitational theories, $\lambda_L$ saturates the chaos bound $\lambda_L\le 2\pi T$ \cite{Maldacena:2015waa}, so they are said to be maximally chaotic. The butterfly velocity, however, depends on the theory \cite{Roberts:2014isa,Mezei:2016wfz,Alishahiha:2016cjk,Qaemmaqami:2017bdn,Li:2017nxh,Dong:2022ucb}.

More recently, it was discovered that the quantities $\lambda_L$ and $v_B$ may already show up in features of the energy density retarded Green's function through a phenomenon called \textit{pole skipping} \cite{Grozdanov:2017ajz,Blake:2017ris,Blake:2018leo}. It was first found numerically for pure Einstein gravity \cite{Grozdanov:2017ajz} and later studied analytically for Einstein gravity with matter \cite{Blake:2018leo}. See also \cite{Grozdanov:2018kkt,Natsuume:2019sfp,Grozdanov:2019uhi,Blake:2019otz,Ahn:2019rnq,Abbasi:2019rhy,Liu:2020yaf,Abbasi:2020ykq,Grozdanov:2020koi,Jansen:2020hfd,Kim:2020url,Jeong:2021zhz,Ageev:2021xjk,Kim:2021hqy,Blake:2021hjj,Mahish:2022xjz,Jeong:2022luo} for holographic and \cite{Haehl:2018izb,Jensen:2019cmr,Das:2019tga,Haehl:2019eae,Ramirez:2020qer,Choi:2020tdj} for boundary studies. 

The retarded Green's function is the relation between a source and its response. Holographically, the Green's function of an operator dual to a bulk dynamic field $X$ (suppressing indices) is given by \cite{Son:2002sd,Iqbal:2009fd}
\begin{equation}
\label{eq:green}
G_{R}\left(\oi,k\right)=\left.\left(\lim _{r \rightarrow \infty} \frac{\left.\Pi\left(r ;\oi,k\right)\right|_{X_{R}}}{X_{R}\left(r;\oi, k\right)}\right)\right|_{X_{0}=0},
\end{equation}
where $X_R$ is the bulk solution satisfying Dirichlet boundary condition $X_R\to X_0$ at infinity and ingoing wave boundary condition at the horizon, and $\Pi$ is its conjugate variable in a radial foliation. 
In terms of an asymptotic expansion, it is proportional to the ratio between the coefficient of the normalizable falloff and that of the non-normalizable falloff. A quasinormal mode, by definition, does not have a non-normalizable divergence, so the poles of the Green's function are identified with the quasinormal spectrum. 


Generically, $X_R$ is uniquely determined from $X_0$, and $G_R$ is therefore well-defined. However, a would-be pole can sometimes get multiplied by a zero, resulting in an ill-defined limit. This happens at a special frequency and momentum, 
\begin{equation}\label{eq:pole}
    \omega = i\lambda_L, \qquad k=\frac{i\lambda_L}{v_B},
\end{equation}
where $\lambda_L$ and $v_B$ are the Lyapunov exponent and the butterfly velocity extracted from a holographic OTOC calculation \eqref{eq:otoc} in Einstein gravity minimally coupled to a large class of matter fields \cite{Grozdanov:2017ajz,Blake:2018leo}.

To explain this universality, \cite{Blake:2018leo} discovered a feature of Einstein's equation at the horizon. Expanding metric perturbations around a stationary planar black hole in terms of Fourier modes, a particular component of Einstein's equation evaluated at the horizon was found to be trivial at \eqref{eq:pole} so that there exists one fewer constraint. This implies an extra degree of freedom of the ingoing modes and consequently an ambiguity in the bulk solution $X_R$ and in turn the Green's function $G_R$. 

Later it was discovered that pole skipping happens more generally at other locations and for other types of Green's functions \cite{Grozdanov:2019uhi,Blake:2019otz,Natsuume:2019xcy,Ceplak:2019ymw,Ahn:2020bks,Ahn:2020baf,Sil:2020jhr,Yuan:2020fvv,Abbasi:2020xli,Ceplak:2021efc,Yuan:2021ets,Kim:2021xdz,Mahish:2022xjz}. See also \cite{Wu:2019esr,Natsuume:2019vcv} for higher-derivative corrections and \cite{Natsuume:2020snz} for a zero-temperature example. However, unlike the one at \eqref{eq:pole}, the other skipped poles are unrelated to chaos.

We put these in the same framework by considering general diffeomorphism invariant bulk theories with matter fields that are not necessarily minimally coupled. For simplicity, we only consider bosonic fields and leave a discussion of fermionic fields to the last section. By defining a weight, we can separate the equations of motion into different groups and evaluate them in a given order. This allows us to find the frequencies of the skipped poles and the corresponding momenta in general. This is done in Section~\ref{sec:pole}. Furthermore, we observe a relation between higher-weight pole-skipping frequencies and higher-spin Lyapunov exponents and use it to justify the removal of a bounded tower of higher spin fields from consideration in the remaining sections. 

In Section~\ref{sec:chaos}, it is shown that, for general higher-derivative gravitational theories, the butterfly velocity can be obtained from the highest-weight equation of motion, and it agrees with the butterfly velocity obtained via a shockwave calculation. This generalizes the matching for Gauss-Bonnet gravity and Einstein gravity with a string theory correction at $O(\alpha'^3)$ \cite{Grozdanov:2018kkt}. We also try to explain this matching between pole skipping and chaos in the same section. By regularizing the metric perturbation at the chaotic skipped pole with a Gaussian distribution in the frequency Fourier space, we obtain a metric that is regular at the horizon. Extending it to a Kruskal–Szekeres coordinate patch and taking the regulator away, we show that this metric perturbation localizes to the past horizon in a distributional sense, like the shockwave metric. We end with a summary and a discussion of potential future directions in Section~\ref{sec:disc}.

\section{\label{sec:pole}General pole-skipping conditions}

The metric for a general stationary planar black hole can be written in ingoing Eddington-Finkelstein coordinates as
\begin{equation}
\label{eq:iEF}
d s^{2}=- f(r) d v^{2}+2 d v d r+h(r) dx^i dx^i,
\end{equation}
where $f(r_0)=0$ at the horizon $r=r_0$ and $i=1,...,d$. The non-vanishing Christoffel components are given by
\begin{equation}\label{eq:chris}
\begin{aligned}
\Gi^{v}_{vv}&=\frac{1}{2}f^\pr,~~~\Gi^{v}_{ij}=-\frac{1}{2}h^\pr\di_{ij},~~~
\Gi^{r}_{vr}=-\frac{1}{2}f^\pr,\\
\Gi^{i}_{rj}&=\frac{h^\pr}{2h}\di^i_j,~~~
\Gi^{r}_{vv}=\frac{1}{2}ff^\pr,~~~\Gi^{r}_{ij}=-\frac{1}{2}fh^\pr\di_{ij}.
\end{aligned}
\end{equation}
For simplicity, we assume that background matter fields are stationary, isotropic and homogeneous in $x^i$, and regular at both past and future horizons, like the metric.

Now, if we define a \textit{pseudo-weight} for any tensor component as the number of lower $v$-indices minus that of lower $r$-indices, where an upper $v$ is considered a lower $r$ and vice versa, then any background tensor component (ones constructed from the stationary background metric and matter fields) with positive weight needs to vanish at the horizon. We prove this next.

In Kruskal–Szekeres coordinates, defined via
\beq
\label{eq:KStransform}
U=-e^{-f^\pr(r_0) (v-2r_*)/2},~~~V=e^{f^\pr(r_0) v/2},
\eeq
where $dr_*/dr=1/f(r)$, one can similarly define a \textit{boost weight} as the number of lower $V$-indices minus that of lower $U$-indices \cite{Wall:2015raa}. Then, the boost symmetry ($V\mapsto a V$, $U\mapsto U/a$) requires that a background quantity with boost weight $n>0$ must scale like $U^{n}$ times a function of the product $UV$, and regularity at the bifurcate horizon requires this function to be non-singular as $UV\to 0$. Therefore, at the future horizon ($U=0$), this vanishes. Relating this to quantities in ingoing Eddington-Finkelstein coordinates, using
\bal
\label{vrtoUV}
d v=\frac{2}{f^\pr(r_0)} \frac{d V}{V},~~~d r=\frac{f(r)}{f'(r_0)}\(\frac{dV}{V}+\frac{dU}{U}\)
\eal
for each lower index $V$ or $U$ of a tensor $T$, we have (suppressing other indices)
\bal
T_{V}=\frac{\pd v}{\pd V}T_v+\frac{\pd r}{\pd V}T_r=\frac{2}{f^\pr(r_0) V}\(T_v+\frac{1}{2} f(r) T_r\)
\eal
and
\bal
T_{U}=\frac{\pd r}{\pd U}T_r=\frac{1}{U}\frac{f(r)}{f^\pr(r_0)} T_r.
\eal
We see that each $V$-index maps to a $v$-index and each $U$-index maps to an $r$-index (all lower indices here but upper ones work similarly) up to terms that are of higher order in $f$. Given that background quantities with positive boost weight and $f$ vanish at the horizon, we arrive at the conclusion that the same is true if we replace boost weight with pseudo-weight. From now on, we no longer need to mention boost weight and will refer to pseudo-weight simply as \textit{weight}.\footnote{The name pseudo-weight emphasizes the fact that it does not correspond to any symmetry transformation, unlike boost weight which characterizes how a tensor component transforms under the boost symmetry. In fact, the boost transformation is just a translation in $v$ in ingoing Eddington-Finkelstein coordinates, and tensor components in this coordinate system do not transform non-trivially under it. The property we need for positive-pseudo-weight quantities is inherited from a more fundamental feature about boost weight via a coordinate transformation and is only true because we can drop terms that vanish at the horizon.}

To describe ingoing quasinormal modes at the horizon, for any dynamic field $X$, we expand its perturbation around the stationary background in the Fourier space as
\begin{equation}
\label{eq:XFourier}
\delta {X}(r, v, x)=\delta {X}(r)\, e^{-i \omega v+i k x}.
\end{equation}
For Einstein gravity, writing Einstein's equation as $E_{\mu\nu}=T_{\mu\nu}$, a particular component under perturbation, $\delta E^r_v$, is proportional to
\begin{equation}\label{eq:einstein}
\left(k^2-i\frac{d}{2} \omega h^{\prime}\right) \delta g_{v v}+( \omega-i2 \pi T)\left[\omega \delta g_{ii}+2 k_i \delta g_{v i}\right].
\end{equation}
On the horizon, for matter perturbations that are regular enough, the stress tensor component $\di T^r_v=0$ \cite{Blake:2018leo}, and prefactors in $\di E^r_v$ can be tuned to zero by choosing \eqref{eq:pole}. As a consequence, Einstein's equation provides one fewer constraint, which serves as an explanation for the universal behaviour of the energy density Green's function with low-spin matter fields coupled to Einstein gravity \cite{Blake:2018leo}. 

Now consider an arbitrary diffeomorphism invariant theory defined with a local action $S=S_g+S_M$ where the gravitational part $S_g$ is given by
\begin{equation}\label{eq:action}
    S_g = \int d^{d+2}x\,\sqrt{-g}\,\mathcal{L}\left(g,R,\nabla,\Phi\right),
\end{equation}
and $S_M$ is part of the action with only minimally coupled matter fields, artificially separated from the rest for later convenience. Here $\mathcal{L}$ can be an arbitrary function of the metric, $g$, and an arbitrary number of bosonic matter fields collectively denoted as $\Phi$. More specifically, $\mathcal{L}$ can be written as a sum of contractions between an arbitrary number of the metric, curvature tensors, matter fields, and an arbitrary number of covariant derivatives of them. 

The metric equation of motion is defined as
\begin{equation}
E_{\mu \nu}=\frac{2}{\sqrt{-g}} \frac{\delta S_g}{\delta g^{\mu \nu}}=-\frac{2}{\sqrt{-g}} \frac{\delta S_M}{\delta g^{\mu \nu}}=T_{\mu\nu}.
\end{equation}
The remaining equations of motion are given by ${\di S}/{\di \Phi}=0$, indices suppressed. Now, to obtain \eqref{eq:green}, the idea is to perturb the dynamical fields and apply the equations of motion everywhere. However, it turns out sufficient to consider the near-horizon expansion of all perturbations in order to study pole skipping. For readability, we introduce the following compact notation: we use $\di \mE=0$ to denote collectively all the perturbed equations of motions and their radial derivatives ($\nabla_r$) evaluated on the horizon. These are essentially the coefficients of a near-horizon Taylor expansion. We further define $\di\mE_p$ as the subset of $\di\mE$ with weight $p$, organized into a vector, and denote its number of components as $|\di\mE_p|$.

Similarly, we collect perturbations of all dynamics fields (including both the metric and matter) and their radial derivatives with weight $q$ into $\di\mathcal{X}_q$ (all evaluated on the horizon). For example, $\di\mathcal{X}_2=(\di g_{vv},\nabla_r \di B_{vvv},...)$ and $\di\mathcal{X}_0=(\di g_{ij},\nabla_r \di g_{vi},\nabla_r\nabla_r \di g_{vv},\di A_i, \nabla_r \di A_v,...)$. 

With these definitions, we can now write
\beq
\di \mE_p= \sum_q {\mM}_{p,q}(\omega,k)\,\di \mathcal{X}_q,
\eeq
where each ${\mM}_{p,q}(\omega,k)$ is a matrix of size $|\di\mE_p|\times |\di\mathcal{X}_q|$. To arrive at this form, first commute all $\nabla_r$'s through $\nabla_i$'s and $\nabla_v$'s to the rightmost location before substituting the Fourier expansion and evaluating the $\nabla_v$'s and $\nabla_i$'s. By definition, the radial derivatives are then absorbed into $\di \mathcal{X}_q$. For later convenience, we also commute all $\nabla_i$ to the right of $\nabla_v$. 

We now prove a useful property that, for $p>q$,
\beq\label{eq:Mpq_factors}
\mM_{p,q}(\omega,k)\propto\[\oi-(p-1)\oi_0\]...\[\oi-q\oi_0\],
\eeq
where $\omega_0=i2\pi T=if'(r_0)/2$.

Begin by noticing that, for a given $p$ and $q'$,
\beq
\label{eq:Fpqkl}
\di\mE_p\sim F(g,R,\nabla,\Phi)(\nabla_v)^k(\nabla_i)^l(\nabla_r)^m\di X_{q'+m}
\eeq
before substituting the Fourier expansion, where $F$ is some $c$-number tensor component constructed out of $g,R,\nabla$ and $\Phi$ such as $R_{v i r j} A^\mu \nabla_\mu\phi$ evaluated on the horizon of the background configuration, and $\di X_{q'+m}$ is the perturbation to some component of a dynamic field $X$ with weight $q'+m$ -- not evaluated on the horizon until acted upon by all the derivative operators in front. Next, notice that the only way to raise weight is with $\nabla_v$ because any background tensor with positive weight vanishes on the horizon. Therefore, to raise the weight of $(\nabla_r)^m\di {X}_{q'+m}$ to that of $\di \mE_p$, one needs $k\ge p-q'$. From \eqref{eq:chris}, it is straightforward to show that, on the horizon,
\beq
\nabla_v T\propto\(\pd_v-\frac{n}{2}f^\pr(r_0)\) T
\eeq
for a general tensor component $T$ with weight $n$;
therefore, evaluating $(\nabla_v)^k$ and substituting \eqref{eq:XFourier} gives at least a factor of $[\oi-(p-1)\oi_0]...[\oi-q'\oi_0]$. Finally, the remaining part $(\nabla_i)^l(\nabla_r)^m\di X_{q'+m}$ evaluates to a number of terms, each proportional to $\di \mX_q$ for some $q\ge q'$. This follows from \eqref{eq:chris}, where any Christoffel symbol appearing in $\nabla_i T$ vanishes if multiplying an object with lower weight than $T$. This concludes our proof of \eqref{eq:Mpq_factors}.

We now discuss the general conditions for pole skipping. We take as an assumption that pole skipping happens whenever an equation of motion becomes trivial.\footnote{It was pointed out in \cite{Blake:2019otz} that there are so-called anomalous points at which triviality of equations of motion at the horizon does not imply dependence on $\di \oi/\di k$ for small deviations from the point, but these points were identified as a different class of skipped poles where the limit does depend on higher order quantities such as $(\di k)^2$ \cite{Ahn:2020baf}. This justifies our assumption here.}  Suppose the highest weight of $\di \mathcal{X}$ is $q_0$, then the highest weight of $\di \mE$ is also $q_0$ (since the action, being a scalar, has weight zero). Consequently, for any positive integer $s$, once we set
\beq
\label{psomega}
\oi=(q_0-s) \, \oi_0,
\eeq
all $\mM_{p,q}(\oi,k)$ with $p>q_0-s\ge q$ are then set to zero (assuming they are not all automatically zero). Now consider the square matrix
\beq
M_s(k)\equiv
\begin{pmatrix}
\mM_{q_0,q_0}&...&\mM_{q_0,q_0-s+1}\\
...&...&...\\
\mM_{q_0-s+1,q_0}&...&\mM_{q_0-s+1,q_0-s+1}\\
\end{pmatrix}
\eeq
where \eqref{psomega} has been substituted. The full set of equations of motion $\di\mE_p, \forall p$, does not determine $\di\mathcal{X}_q, \forall q$, when
\beq
\label{psk}
{\rm det}~M_s(k)=0.
\eeq
The equations \eqref{psomega} and \eqref{psk} are therefore the generalized pole-skipping conditions (for any given $s\ge1$), assuming the second one has solutions. If the theory has a highest spin field with bounded spin $l$, then $q_0=l$ and the pole-skipping frequencies are $(l-s)\,\oi_0$, consistent with observations made in \cite{Blake:2019otz,Haehl:2018izb,Kim:2021xdz} and in particular reproducing the positions of pole skipping at Matsubara frequencies first found in \cite{Blake:2019otz}. The second condition is a polynomial equation for $k$, and the roots are then the pole-skipping momenta, which could be more than one. The order of the polynomial increases with the size of the matrix, and therefore there will be generically more pole-skipping points at larger $s$ (lower $\oi$). 

The first pole skipping happens at $s=1$ at frequency $\omega=(q_0-1)\,\omega_0=i(q_0-1)2\pi T$. Suppose there exists an equation of motion with e.g. three lower $v$-indices. In that case, there will be a skipped pole at $2\omega_0=i4\pi T$, and the field perturbation \eqref{eq:XFourier} will grow like $\exp({4\pi T t})$. On this ground, we expect (finitely many) higher spin fields to violate the chaos bound. This is supported by an independent calculation of the spin-$l$ Lyapunov exponent, $\lambda_L^l=(l-1)2\pi T$ \cite{Perlmutter:2016pkf} and is consistent with the findings of \cite{Haehl:2018izb,Narayan:2019ove}. Bounded higher spin fields also suffer from causality violation \cite{Camanho:2014apa}, which is another reason to exclude them from consideration in the next section. 
Notice, however, that equations of motion for fields with no dynamics automatically have $\mM_{p,q}(\oi,k)=0$ for $p>q$ due to the nonappearance of $\nabla_v$, so they do not become trivial from non-trivial; therefore, they do not violate the chaos bound, in agreement with \cite{Perlmutter:2016pkf} where pure AdS${}_3$ higher spin gravity was exempt from their argument for bound violation.

If $q_0=2$, which is the case for an arbitrary metric theory coupled to matter fields of spin no larger than two, then the bound is satisfied and in fact saturated. We will discuss this further in the next section. 


For $q_0<2$, such as a scalar or vector field without gravitational backreaction, there is no growing mode and therefore no relation to chaos, but an infinite number of skipped poles still exist and constrain the structure of Green's functions \cite{Grozdanov:2019uhi,Blake:2019otz}. 

\section{\label{sec:chaos}Matching of butterfly velocities}
For arbitrary higher-derivative gravity coupled to scalar, vector or form fields, $q_0=2$ (from the metric) and the highest-weight skipped pole has $\omega=i2\pi T$. We now show that the corresponding butterfly velocity matches that obtained from the OTOC.

In this case, the only dynamic field with weight 2 is $\di\mathcal{X}_2= \di g_{vv}$, and the corresponding equation of motion is $\di\mE_2=\di E_{vv}-\di T_{vv}=0$. The perturbation to the stress tensor component $\di T_{vv}$ does not necessarily vanish, but $\di T^r_v$ ($=\di T_{vv}-T_{rv}\di g_{vv}$) does vanish for matter fields regular on the horizon \cite{Blake:2018leo}. We will make this restriction in order to compare results with OTOC: the metric shockwave also has vanishing $\di T^r_v$. Therefore, the pole-skipping conditions with $s=1$ are given by 
\beq
\oi=\oi_0,~~~{\rm det}~M_1=\frac{\di E^r_v}{\di g_{vv}}=0.
\eeq
This gives a polynomial equation for $k$ with only even powers (by symmetry). In cases where the polynomial is of quartic order or higher, one can take the view that all corrections to Einstein gravity should be treated perturbatively so only the roots continuously connected to Einstein gravity are physical. But as we will see, the matching is evident without a perturbative treatment.

For the class of theories we consider,
\beq
\di E^r_{v}=\sum_{k,l} H_{k,l}(f,h,\pd_r,\Phi)(\pd_v)^{k}(\pd_i)^{l} \di g_{vv}
\label{eq:Erv}
\eeq
for some non-covariant $c$-number coefficients $H_{k,l}$. The non-trivial statement that no $\pd_r$ acts on $\di g_{vv}$ and none of the other components such as $\di g_{vi}$ can appear follow directly from the weight argument. As an example, consider the Einstein gravity equation of motion \eqref{eq:einstein} studied in \cite{Blake:2018leo}. Since $\di g_{ij}$ has weight zero, it has to pick up a factor of $\oi$ to get to weight one and then a factor of $(\oi-\oi_0)$ to get to weight two, similarly for $\di g_{vi}$ which only needs to raise its weight by one. Another simplification in Einstein gravity is due to the fact of it being two-derivative. It is not possible for \eqref{eq:einstein} to contain a term like for example $\pd_r\delta g_{vi}$: this quantity has weight zero and therefore needs two $v$-derivatives to go to two, but it already has one derivative itself. 

To compare this with the shockwave calculation, we move to Kruskal–Szekeres coordinates defined in \eqref{eq:KStransform}. Then $UV=-e^{f^\pr(r_0) r_*}$, and the metric is given by
\beq
\d s^2=2A(UV) \d U \d V+B(UV) dx^i dx^i,
\eeq
\beq
A(UV)=\frac{2}{f^\pr(r_0)^2}\frac{f(r)}{UV} ,~~~B(UV)=h(r).
\eeq

In general higher-derivative gravity and for a shockwave along $V=0$, the only non-trivial component of $\di E^\mu_\nu$ perturbed by a local source is $\delta E^U_V$ \cite{Dong:2022ucb}. For a general perturbation, $\di g_{vv}$, translating to ingoing Eddington-Finkelstein coordinates, this component is given by
\beq
\di E^{U}_{V}=\frac{U}{V}\(\frac{2}{f(r)} \di E^r_v+\di E^r_r-\di E^v_v-\frac{f(r)}{2}\di E^v_r\).
\label{eq:map}
\eeq
Compared to the first term, others are suppressed with extra factors of $f(r)$, so they vanish when evaluated on the horizon. Similarly, $\di T^U_V\propto \di T^r_v$, but recall that this vanishes for regular matter configurations. Therefore,
\bal
\label{eq:big_EUV}
0&=\di E^{U}_{V}=\frac{2UV}{f(r)} \frac{1}{V^2} \sum_{k,l} H_{k,l}(\pd_v)^{k}(\pd_i)^{l} \di g_{vv}\nn
&=\frac{2UV}{f(r)} \frac{1}{V^2} \sum_{k,l} H_{k,l}(\pd_i)^{l}\(\frac{2}{f^\pr(r_0)}V\pd_V\)^{k} \di g_{vv}\nn
&=\frac{2UV}{f(r)} \sum_{k,l} H_{k,l}(\pd_i)^{l}\(\frac{2}{f^\pr(r_0)}(V\pd_V+2)\)^{k}\frac{ \di g_{vv} }{V^2}\nn
&=\frac{2UV}{f(r)} \sum_{k,l} \tilde{H}_{k,l}(\pd_i)^{l}\(V\pd_V\)^{k}\frac{ \di g_{vv} }{V^2}\nn
&=\frac{1}{A} \sum_{k,l} \tilde{H}_{k,l}(\pd_i)^{l}\(V\pd_V\)^{k} \di g_{VV},
\eal
where we used the transformation $\pd_v=\frac{2}{f^\pr(r_0)}V\pd_V$ in going to the second line, and a trick
\beq
\frac{1}{V^2} V\pd_V =\(V\pd_V+2\) \frac{1}{V^2}
\eeq
in going to the third line. The fourth line follows from a reorganization of the sum with new coefficients $\tilde H_{k,l}$ and the last line follows from
\begin{equation}
\label{eq:gvv_gVV}
    \di g_{VV}(V,x)=\frac{4\di g_{vv}(v,x)}{f^\pr(r_0)^2 V^2}.
\end{equation}

The special thing about $\omega=\omega_0$ is that,
\bal
\di g_{vv}\sim e^{-i\oi_0 v}= e^{-\frac{i2\oi_0}{f^\pr(r_0)} \log V}=V,
\eal
and therefore
\begin{equation}\label{eq:dgVV_final}
    \di g_{VV}(V,x)\sim \frac{1}{V}\,e^{-ikx}.
\end{equation}
Compare this with a linearized shockwave perturbation
\beq
\label{eq:shockmet}
\di g_{VV} \sim \di(V)\,e^{-\mu{x}},
\eeq
where $\mu=2\pi T/v_B$ upon using $\di E^U_V=0$ (outside of a localized source in $x$). Noticing that $\di (V)$ has the same distributional behavior as $1/V$ under $V\pd_V$ \cite{Grozdanov:2017ajz}, e.g., $V\di' (V)=-\di(V)$ and $V d(1/V)/dV=-1/V$, it follows that $k=i2\pi T/v_B$ upon using \eqref{eq:big_EUV} for the perturbation \eqref{eq:dgVV_final}, thereby extending \eqref{eq:pole} to general higher-derivative gravity and hence some of the results of \cite{Blake:2018leo,Grozdanov:2018kkt}.

Given the similarity between $1/V$ and $\di(V)$ and the role this similarity plays in establishing the equivalence of these two calculations of the butterfly velocity, it is natural to wonder whether there is a more direct connection between them. An immediate obstacle is the divergence of the function $1/V$ at the past horizon $V=0$. We mitigate this problem with an unnormalized\footnote{We should note that the need for the unnormalized regulator arises from the need to remove the divergence. Alternatively, one can use a normalized delta function regulator and remove the divergence at the end using a subtraction not unlike the minimal subtraction in dimensional regularization.} regularization of the Fourier space delta function along the real frequency line:
\begin{equation}
    \int d\xi \,\delta (\xi) \to \int_{-\infty}^\infty \,{d\xi} \,e^{- {\xi^2}/{a}} ,
\end{equation}
giving rise to a mode
\beq \label{eq:gaussian}
 {\di g_{vv}}={\sqrt{\pi a}}\, e^{-\frac{a}{4}(\li_L v)^2}e^{\li_L v}.
\eeq
To compare with the shockwave metric \eqref{eq:shockmet}, we convert this to Kruskal-Szekeres coordinates. Using \eqref{eq:gvv_gVV},
\beq
\di g_{VV}=
\begin{cases}
0,&V<0 \\
\frac{\sqrt{\pi a}}{\lambda_L^2 }\frac{1}{V}\, e^{-a(\log V)^2/4}, &V\ge0
\end{cases}
\eeq
where we have used the fact that the perturbation vanishes exactly behind the past horizon. This function is finite and integrates to a constant for finite $a$, and it vanishes everywhere off the horizon as $a\to0$. It therefore behaves as a regularized $\di (V)$. Taking the regulator away, this becomes a shockwave localized at $V=0$.\footnote{Physically, this suggests that the shockwave solution encodes part of the physical content of the quasinormal mode at the highest-weight pole-skipping frequency. The renormalization procedure throws away some information irrelevant for computing the butterfly velocity.}


\section{\label{sec:disc}Discussion}
We have defined a quantity called weight to organize bulk equations of motion and exploited its convenience to show that pole skipping happens in holographic CFTs dual to quite general diffeomorphism invariant bulk theories. As a result, the pole-skipping frequencies show up at $(q_0-s)\,\omega_0$ for all $s\in \mathbb{Z}^+$, where $\oi_0=i2\pi T$, and $q_0$ is defined as the weight of the highest-weight object. In particular, a theory that has a bounded highest spin larger than two in general gives rise to $q_0>2$, which leads to very fast scrambling that violates the chaos bound. It is therefore reasonable to disallow a finite tower of higher spin fields, in addition to causality reasons \cite{Camanho:2014apa}. This brings down $q_0$ to two, and, with this restriction, the metric is the field that can have the highest weight. This is the main reason behind the universality of the special pole-skipping point at $\omega=i\lambda_L$ and $k=i\lambda_L/v_B$, where $\lambda_L=2\pi T$, and $v_B$ is defined via a OTOC calculation. 

In other words, for maximally chaotic holographic theories, instead of needing to compute a four-point function, the retarded Green's function already knows about the butterfly velocity, and its dependence on the bulk theory is exactly the same as an OTOC would predict. It would be interesting to test this statement for non-holographic maximally chaotic theories.\footnote{For non-maximally chaotic theories, the predictions from pole skipping could differ from OTOC results \cite{Mezei:2019dfv,Choi:2020tdj,Blake:2021wqj}.} Furthermore, there are now three ways of computing the butterfly velocity: (i) using entanglement wedge, (ii) using shockwave and (iii) using pole skipping. We proved the equivalence between the second and third \textit{prescriptions} themselves.\footnote{Evidence for the general equivalence between the first two was presented in \cite{Mezei:2016wfz,Dong:2022ucb}; evidence for the general equivalence between the last two was presented in \cite{Blake:2018leo,Grozdanov:2018kkt}. Here our emphasis is on the equivalence of the methods and not the equality of the results.}

The restriction of the discussion to bosons is for simplicity, and the generalization to include fermions should be completely analogous. For minimally coupled spinors on a fixed background, pole skipping has been shown to happen at $\oi=(q_0-s)\,\oi_0$ for a half integer $q_0=1/2$ and positive integers $s$ \cite{Ceplak:2019ymw}; with a spin-$3/2$ Rarita-Schwinger field, $q_0$ becomes $3/2$ \cite{Ceplak:2021efc}. Both of the examples fit the pattern that the leading pole skipping happens at $(q_0-1)\,\oi_0$, and if one allows both bosonic and fermionic fields with arbitrary couplings between them, one might expect that both $q_0$ and $s$ can be half integers. It might be of use to analyze this with the weight argument, perhaps beginning by rephrasing the current discussion in a spin connection language. 


We should summarize three assumptions that were used: (i) the existence of a finite $q_0$; (ii) the non-triviality of equation \eqref{psomega}, i.e., the entries set to zero by this equations are not already all zero; and (iii) equation \eqref{psk} has solutions. We expect that assumption i can be lifted with more careful analysis, but assumptions ii and iii are essential. Given any theory, one needs to check whether these are satisfied. For example, Vasiliev gravity violates assumption i as it contains an infinite tower of higher spin fields; this is consistent with it being dual to a sector of a free theory \cite{Klebanov:2002ja}, which does not exhibit chaos. 

Another condition implicit in our discussion is the restriction to finite temperatures. Extremal black holes do not have a bifurcate surface, so the property derived from regularity at the bifurcate surface no longer applies. Furthermore, poles in the Green's function get replaced by branch cuts \cite{Faulkner:2009wj,Natsuume:2020snz}. Accordingly, a generalization of our argument to zero temperature will be non-trivial.

We also showed that the shockwave metric could be obtained from a regularized mode of the metric perturbation. This serves as an explanation for the similarities between the two calculations and the equivalence regardless of the theory. One might try different regulators or use different subtraction schemes to find a more regulator-independent relation. 

\noindent\textbf{Note added in v2:} Outside the chaos bound, the relation between pole skipping and the OTOC is not well understood, and it is unclear whether the leading pole-skipping frequency can be identified with the Lyapunov exponent. In the current version, we only make the observation that they are equal and leave its investigation to the future. We thank Mike Blake for pointing this out.

\begin{acknowledgments}
We thank Xi Dong, Gary Horowitz, Mark Mezei, Aron Wall, Zihan Yan and especially Mike Blake for valuable feedback. DW thanks the organizers of the conference ``Fundamental Aspects of Gravity" at Imperial College London, August 2022, for the invitation to present this work, the attendees for enlightening discussions, and the ``Gravity Theory Trust" for its support. DW is supported by NSF grant PHY2107939.

\end{acknowledgments}


\bibliography{bibliographyPRL}

\end{document}